\documentclass[aip,jmp, amsmath,amssymb,floatfix,reprint]{revtex4-1}
\usepackage{graphicx}
\usepackage{dcolumn}
\usepackage{bm}

\begin{document}


\title{Fabrication of sub-15nm aluminum wires by controlled etching} 



\author{T. Morgan-Wall}
\author{H. J. Hughes}
\author{N. Hartman}
\affiliation{ 
Department of Physics and Astronomy, Johns Hopkins University, Baltimore, Maryland 21218, USA.
}
\author{T. M. McQueen}
\affiliation{ 
Department of Physics and Astronomy, Johns Hopkins University, Baltimore, Maryland 21218, USA.
}
\affiliation{ 
Department of Chemistry, Johns Hopkins University, Baltimore, Maryland 21218, USA.
}
\author{N. Markovi\'{c}}
\affiliation{ 
Department of Physics and Astronomy, Johns Hopkins University, Baltimore, Maryland 21218, USA.
}



\begin{abstract}
We describe a method for fabrication of uniform aluminum nanowires with diameters below 15 nm. Electron beam lithography is used to define narrow wires, which are then etched using a sodium bicarbonate solution, while their resistance is simultaneously measured in-situ. The etching process can be stopped when the desired resistance is reached, and can be restarted at a later time. The resulting nanowires show a superconducting transition as a function of temperature and magnetic field that is consistent with their diameter. The width of the transition is similar to that of the lithographically defined wires, indicating that the etching process is uniform and that the wires are undamaged. This technique allows for precise control over the normal state resistance and can be used to create a variety of aluminum nanodevices. 
\end{abstract}


\maketitle 


Superconducting properties of a nanowire become significantly altered as its diameter is decreased to the order of the coherence length \cite{Oreg}. The superconducting transition itself becomes broadened \cite{Lau, Markovic} due to the phase slips, which can occur though thermal activation \cite{LA,MH} or quantum tunneling \cite{Golubev}. As the temperature is further decreased towards absolute zero, one-dimensional nanowires are expected to undergo a quantum phase transition \cite{Sachdev, Refael, Herbut, DelMaestro} between a superconducting and an insulating state. The nature of this transition generally depends on disorder, magnetic field, boundary conditions and the coupling to the environment. Superconducting nanowires could be a useful model system in which to study these effects, but in order to do a systematic study, one needs a variety of nanowires with well controlled properties. 
Various methods have been used to fabricate superconducting nanowires with dimensions that are smaller than what is achievable by electron beam lithography: step edge method \cite{Giordano}, molecular templating \cite{Bezryadin2,Bezryadin}, electrodeposition into nanopores \cite{Chan,Michotte}, mechanical stencil methods \cite{natelson,altomare}, and focused ion beam milling \cite{Zgirski}.
All these methods have succeeded in reducing the diameters of the nanowires and have enabled discovery of new phenomena, but each has its limitations and disadvantages. For example, ion milling will typically damage the samples to some extent by ion implantation, molecular templating offers very little control over the size and the structure of the wires, and the electrodeposition in nanopores yields nanowires that usually remain encapsulated in the nanopores. Additionally, most of these methods do not lend themselves to four-probe measurements, nor offer good control of the dissipative environment. 
In this work, we describe a controlled etching method for fabrication of uniform aluminum nanowires with diameters below 15nm. The resistance is measured \textit{in situ} as the wire is being etched, allowing a precise control of normal state resistance of each wire. Given that aluminum 
is often the material of choice in superconducting devices, this method offers a clean and controllable alternative for fabrication of superconducting nanostructures on a variety of substrates.

\begin{figure}
\includegraphics[width=\columnwidth]{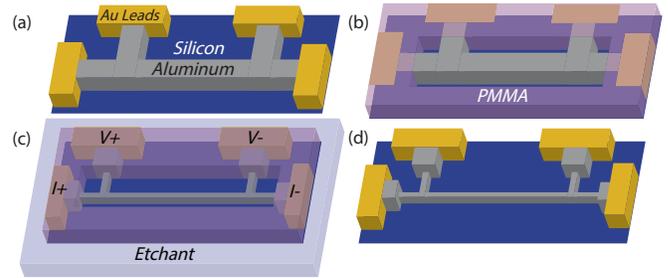}
\caption{\label{Figure 1} (a) An aluminum nanowire with a four-probe lead configuration is created using electron beam lithography and thermal evaporation of 80-100nm of aluminum on top of pre-fabricated gold leads. The gold leads were fabricated using optical lithography and thermal evaporation of 5nm of Cr, followed by 80nm of Au. (b) The gold leads are wirebonded to a chip carrier, and a PMMA etching mask is created using electron beam lithography. (c) As a drop of the etching solution is placed on the sample, the area defined by the window in the PMMA mask is etched, while the resistance of the sample is monitored in a four-probe configuration. (d) When the desired resistance is reached, the etching is stopped by removing the droplet of the etchant using $N_2$ gas. The PMMA is then dissolved in acetone, leaving only the etched wire.}
\end{figure}

We start with clean silicon substrates, on which we define gold bonding pads using standard ultraviolet lithography and thermal evaporation. The substrates are then cleaned and spin-coated with 4\% poly-methyl-methacrylate (PMMA) at 4000 RPM, patterned using electron beam lithography (EBL), and developed using the cold methyl isobutyl ketone (MIBK) process \cite{coldmibk}. 80-100nm of aluminum is thermally evaporated onto the substrate in a vacuum of $10^{-7}$ Torr. After evaporation, the nanowires are sonicated in acetone to achieve uniform lift-off. Figure 1 shows a schematic 
of a nanowire before, during and after the etching process. The nanowires are patterned in a four-probe configuration, as shown in Fig 1.a., with 
dimensions of 200nm in width and 5$\mu$m in length, as measured between the voltage leads. The samples are then wirebonded to the chip carrier 
using 25$\mu$m wide aluminum wires. After wirebonding, a layer of 4\% PMMA is spun at 4000 RPM on top of the sample, and a window is 
created using e-beam lithography, as shown in Fig. 1 b. The window defines the area that is to be etched, while the PMMA protects the rest of 
the sample, serving as the etching mask. 

At this point, the sample is connected to a lock-in amplifier (Princeton Applied Research 124A) and the resistance is measured in a four-probe configuration (typically using 100nA at 27Hz). To begin the etching process, a droplet of the solution can now be placed on top of the sample. 
The etching solution was 100mL deionized water : 5g sodium bicarbonate : less than 50mg sodium nitrite. The concentration of the sodium bicarbonate was chosen in order to set the pH between the barrier of the corrosion and passivation zones in the Pourbaix diagram for aluminum \cite{pourbaix}. The trace amount of sodium nitrite acts to prevent over-oxidation of the aluminum, which would result in insulating nanowires. As the area defined by the PMMA mask is etched, as shown in Fig 1.c., the resistance is monitored as a function of time. When the desired resistance is reached, the solution is removed using compressed $N_2$, and the PMMA is dissolved in acetone, leaving a thinner, narrower nanowire (Fig. 1.d).

\begin{figure}
\includegraphics[width=\columnwidth]{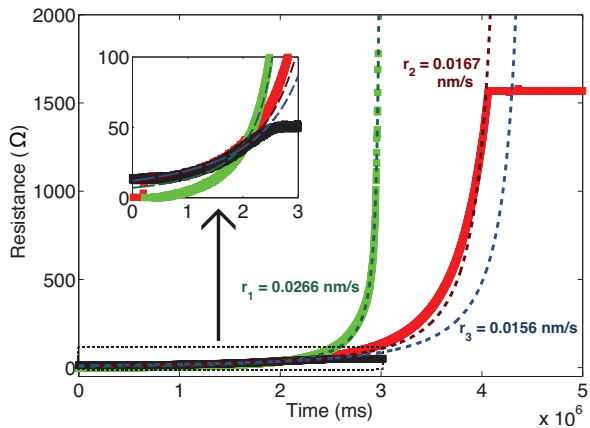}
\caption{\label{Figure 2} Resistance as a function of time for three aluminum nanowires (solid lines), measured during and after the etching process. Nanowire 1 (green) was etched all the way through, and its resistance diverged when the wire was no longer continuous. Nanowire 2 (red) was etched until its resistance reached 1600$\Omega$, at which point the etching solution was abruptly removed, and nanowire 3 (black) was etched until the resistance reached 50$\Omega$. The inset zooms in the lower resistance region to show that the resistance of nanowire 3 remains at 50$\Omega$ after the etching is stopped. The expression in Eq. 1 was used to fit the data, using the etching rate as the fitting parameter. The fits are shown as dashed lines, with the extracted etching rates shown for each nanowire. All three nanowires were etched at the ambient temperaure of 25C.}
\end{figure}

The resistance as a function of time for three representative nanowires is shown in Fig. 2. One of these nanowires was etched all the way through, to the point where it was no longer electrically connected (shown in green in Fig. 2). Another nanowire was etched until the resistance reached  1600$\Omega$, at which point the etching process was stopped by abruptly blowing the solution off (shown in red in Fig. 2). After the etching solution was removed, the resistance of the nanowire remained constant at 1600$\Omega$. The third nanowire (shown in black in Fig. 2) was etched until the resistance reached 50$\Omega$, at which point the droplet of the etching solution was allowed to gradually slide off the sample. The resistance also remained firmly at 50$\Omega$, as shown in the inset of Fig. 2. The minimum size (and thus maximum resistance) of a nanowire made by this technique is controlled by the initial dimensions of the wire. Nanowires with an initial width-to-thickness ratio of 2:1 are found to retain that ratio as they are etched, while nanowires with other width-to thickness ratios are etched entirely through the smallest dimension first.

\begin{figure}
\includegraphics[width=\columnwidth]{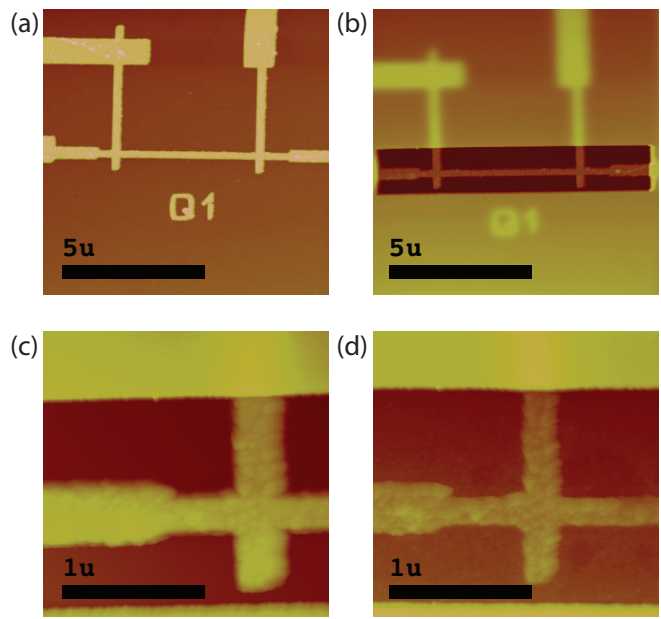}
\caption{\label{Figure 3} (a) Atomic force microscope (AFM) image of an aluminum wire before the PMMA mask is placed on top. (b) AFM image with the PMMA window. (c) A close-up AFM image of a section of a nanowire before etching. (d) A close-up AFM image of the same section of the nanowire shown in (c) after etching for 600 seconds. Note that the nanowire is both thinner and narrower, and the measured dimensions agree with the etching rate extracted from the resistance measurements. }
\end{figure}

Assuming that the nanowires are uniform enough that the resistivity is constant, we construct a simple model to describe the etching process. 
Since the dimensions of the nanowire are changing as a function of time, the resistance of the nanowire will be determined by the changes in 
its thickness and the width in the following way:
\begin{equation}
R_{\text{wire}}(t)=\frac{\rho L}{(W - 2 r t)(T -  r t)}
\end{equation}
here $\rho$ is the resistivity of aluminum, L is the length of the nanowire, T is the thickness, W is the width, r is the etching rate and t is 
time. This model assumes that the nanowire has a rectangular cross-section and that the etching rate is the same on each face. The width (W) decreases at twice the rate as the thickness (T), because the nanowire is etched from both sides and from the top, but not from the bottom.

Using the measured values for L, T and W, and the resistivity for bulk aluminum, the etching rate can be treated as a fitting parameter and 
extracted from the data in Fig. 2. The fits to the resistance as a function of temperature and the corresponding etching rates are shown in 
Fig. 2. In order to account for slight variations in the width due to the resolution limits of our lithography, we introduced a second fitting parameter for the width. This parameter turned out to be of order one, which means the starting width of the nanowire was close to 
the intended design. The etching rates were found to be in the range of 0.01-0.03 nm/s for the chosen concentrations of sodium bicarbonate in the etching solution. The etching rate can be controlled by adjusting the concentration of the sodium bicarbonate in the etching solution - a lower
pH factor resulted in a slower etching rate. The etching rate was also somewhat sensitive to the ambient temperature, and could be increased by placing the sample under a heat lamp (the results shown in Fig. 2. were obtained at the ambient temperature of 25C).

The etching rate and the uniformity of the etched nanowires is further confirmed by atomic force microscopy (AFM). An AFM image of 
a nanowire before etching is shown in Fig. 3.a (illustrated in the schematic in Fig. 1.a), and Fig. 3.b. shows a nanowire with 
the PMMA mask on top (as illustrated in Fig. 1.b). A close-up of a section of a nanowire is shown before etching (Fig. 3.c) and 
after etching (Fig. 3.d). It is evident that the wire is both thinner (indicated by the darker color on the image) and narrower. 
The etching rate can be determined from the height measurements, and it is consistent with the etching rates extracted from the resistance measurements. 

\begin{figure}
\includegraphics[width=\columnwidth]{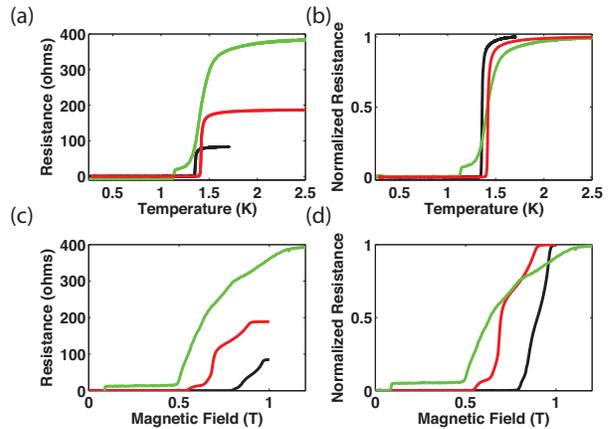}
\caption{\label{Figure 4} (a) Resistance as a function of temperature of a nanowire (with $R_{N}$=400$\Omega$) produced using the etching method, compared with two nanowires ($R_{N}$=80$\Omega$ and $R_{N}$=180$\Omega$) produced 
using electron beam lithography. (b) Normalized resistance as a function of temperature for the same nanowires. The etched nanowire has a wider 
transition than the lithographically produced wires, which is expected for a nanowire with a smaller cross-section. (c) Resistance as function of magnetic field for the same nanowires. The transition broadens as the nanowire gets more resistive. (d) Normalized resistance as a function 
of magnetic field. The non-zero resistance at the bottom of the etched nanowire's transition occurs because this particular nanowire was etched 
only between the voltage leads and not over them, so the measurement includes a small non-etched section that goes through the transition 
at a different magnetic field.}
\end{figure}

Superconducting properties of the etched nanowires were measured down to 250mK in a He3 cryostat. Resistance as a function of temperature for an etched nanowire with a normal state resistance $R_{N}$=400$\Omega$ and two lithographically produced nanowires with $R_{N}$=80$\Omega$ and $R_{N}$=180$\Omega$ is shown in Fig. 4. a. The etched nanowire undergoes the superconducting transition around 1.4K, at a similar temperature as the lithographically produced nanowires (additional transition at 1.2K is closer to the bulk transition temperature for aluminum, and it is due to a small unetched section between the voltage leads on that particular sample). The smoothness of the transition further confirms that 
the nanowire is fairly uniform, without large variations in width and thickness. In order to compare the transition widths for the three 
nanowires, we show their normalized 
resistance as a function of temperature in Fig. 4.b. The etched nanowire exhibits a slightly broader transition, which is typically found in nanowires 
that have a smaller diameter \cite{Lau}. Resistance as a function of magnetic field is shown in Fig. 4.c., with the normalized resistance 
shown in Fig. 4.d. The transition to the normal state for the etched nanowire occurs at a slightly lower magnetic field, and the transition is broader and smoother than the transitions of the lithographically defined nanowires. Both the temperature and the magnetic field dependence 
of the resistance are consistent with the AFM images, which show that the etched nanowire is uniform and has a smaller cross section than 
the lithographically defined nanowires.

The controlled etching method allows for creating uniform aluminum nanowires with a precise geometry and the normal state resistance,
but it is not limited to making nanowires. The geometry of the etched area is determined by the PMMA mask, which means that arbitrary 
areas can be etched while others are kept intact. Combined with the real-time monitoring of the resistance of the device, the slow etching
allows for great control over the final sample's dimensions.


%
%

%

\begin{acknowledgments}
This work was supported by NSF DMR-1106167. N. M.
would like to thank the Aspen Center for Physics for 
hospitality. T.M.M. acknowledges support from the David and Lucile Packard Foundation.
\end{acknowledgments}

%

\end{document}